%% file: ausdm.tex
\documentclass[english]{article}
\usepackage[latin9]{inputenc}
\usepackage{geometry}
\usepackage[authoryear]{natbib}
\renewcommand{\cite}{\citep}
\usepackage{amsmath,amssymb,amsfonts}
\usepackage{times}
\usepackage{epsfig}
\usepackage{amsmath}
\usepackage{amssymb}
\usepackage{amsthm}
\usepackage{algorithm}
\usepackage{graphicx}
\newcommand{\comment}[1]{}
\usepackage{epstopdf}

\newcommand{\U}{\mathcal{U}}
\newcommand{\I}{\mathcal{I}}
\newcommand{\M}{\mathcal{M}}

\begin{document}

\title{Preference Networks: Probabilistic Models for Recommendation Systems}
\author{Tran The Truyen, Dinh Q. Phung and Svetha Venkatesh \\
Department of Computing, Curtin University of Technology \\
GPO Box U 1987, Perth, WA, Australia}

\date{}

\maketitle

\begin{abstract}
Recommender systems are important to help users select relevant
and personalised information over massive amounts of data
available. We propose an unified framework called
Preference Network (PN) that  jointly models 
various types of domain knowledge for the task of recommendation. 
The PN is a probabilistic model that systematically combines both content-based filtering
and collaborative filtering into a single conditional Markov random field.
Once estimated, it serves as a probabilistic database that supports
various useful queries such as rating prediction and top-$N$ recommendation.
To handle the challenging problem of learning large networks of users and items,
we employ a simple but effective pseudo-likelihood with regularisation. 
Experiments on the movie rating data demonstrate the merits of the PN.
\end{abstract}
\vspace{.1in}

\noindent {\em Keywords:} Hybrid Recommender Systems, Collaborative Filtering,
							Preference Networks, Conditional Markov Networks, Movie Rating.

\section{Introduction}
With the explosive growth of the Internet, users are currently 
overloaded by massive amount of media, data and services. Thus 
selective delivery that matches personal needs is very critical.
Automated recommender systems have been designed for this purpose, and
they are deployed in major online stores such as
Amazon~[http://www.amazon.com], Netflix~[http://www.netfix.com] and new services such
as Google News~[http://news.google.com].

Two most common tasks in recommender systems are 
predicting the score the user might give for a product 
(the \emph{rating prediction task}), and recommending
a ranked list of most relevant items (the \emph{top-$N$ recommendation task}).
The recommendations are made on the basis of the content
of products and services (\emph{content-based}), or based on collective preferences of
the crowd (\emph{collaborative filtering}), or both (\emph{hybrid methods}).
Typically, content-based methods work by matching product attributes
to user-profiles using classification techniques. 
Collaborative filtering, on the other hand, relies on
preferences over a set products that a given user and others have expressed. 
From the preferences, typically in term of
numerical ratings, \emph{correlation-based methods}
measure similarities between users \cite{resnick94grouplens} (user-based methods) 
and products \cite{sarwar2001ibc} (item-based methods). 
As content and preferences are complementary,
hybrid methods often work best when both types of information
is available \cite{balabanovic1997fcb,basu1998rcu,pazzani1999fcc,schein2002mam,basilico2004uca}.

\emph{Probabilistic modeling} \cite{breese1998eap,heckerman2001dni,hofmann2004lsm,marlin2004mur} 
has been applied to the recommendation problem 
to some degree and their success has been mixed. 
Generally, they build probabilistic models that explain data.
Earlier methods include
Bayesian networks and dependency networks \cite{breese1998eap,heckerman2001dni}
have yet to prove competitive against well-known correlation-based counterparts.
The more recent work attempts to perform
clustering. Some representative
techniques are mixture models, probabilistic latent semantic analysis (pLSA)
\cite{hofmann2004lsm} and latent Dirichlet allocation (LDA)
\cite{marlin2004mur}. These methods are \emph{generative} in the sense
that it assumes some hidden process that generates
observed data such as items, users and ratings. The generative assumption
is often made for algorithmic convenience and but it does not necessarily
reflect the true process of the real data.

\emph{Machine learning} techniques~\cite{billsus1998lci,basu1998rcu,basilico2004uca}
address the rating prediction directly without making
the generative assumption. Rather, they map the recommendation into a classification problem
that existing classifiers can solve \cite{basu1998rcu,zhang2002rsu}. 
The map typically considers each user or each item as an
independent problem, and ratings are training instances.
However, the assumption that training instances 
are independently generated does not hold in collaborative filtering. Rather all the ratings
are interconnected directly or indirectly 
through common users and items. 

To sum up, it is desirable to build a recommendation system that
can seamlessly integrate content and correlation information
in a disciplined manner. At the same time, the system should address
the prediction and recommendation tasks directly without replying
on strong prior assumptions such as generative process and 
independence. To that end, we propose a probabilistic graphical formulation
called Preference Network (PN) that has these desirable properties.
The PN is a graph whose vertexes represent ratings (or preferences) and edges
represent dependencies between ratings. The networked ratings are treated as
random variables of conditional Markov random fields~\cite{lafferty01conditional}.
Thus the PN is a formal and expressive formulation that supports
learning from existing data and various inference tasks
to make future prediction and recommendation. 
The probabilistic dependencies between ratings capture the
correlations between co-rating users (as used in~\cite{resnick94grouplens}) 
and between co-rated items (as used in~\cite{sarwar2001ibc}). 

Different from previous probabilistic models, the PN does not
make any generative assumption. Rather, prediction of preferences is
addressed directly based on the content and prior ratings available
in the database. It also avoids the independence assumption made
in the standard machine learning approach by supporting \emph{collective classification}
of preferences. The nature of graphical modeling
enables PN to support missing ratings and joint
predictions for a set of items and users. It
provides some measure of \emph{confidence} in each prediction made,
making it easy to assess the nature of recommendation
and rank results. More importantly, 
our experiments show that the PNs
are competitive against the well-known user-based method
\cite{resnick94grouplens} and the item-based method~\cite{sarwar2001ibc}.


\section{Recommender Systems}
\label{sec:recom-sys}

\begin{figure}[htb]
\begin{center}
\begin{tabular}{ccc}
\includegraphics[width=0.45\linewidth]{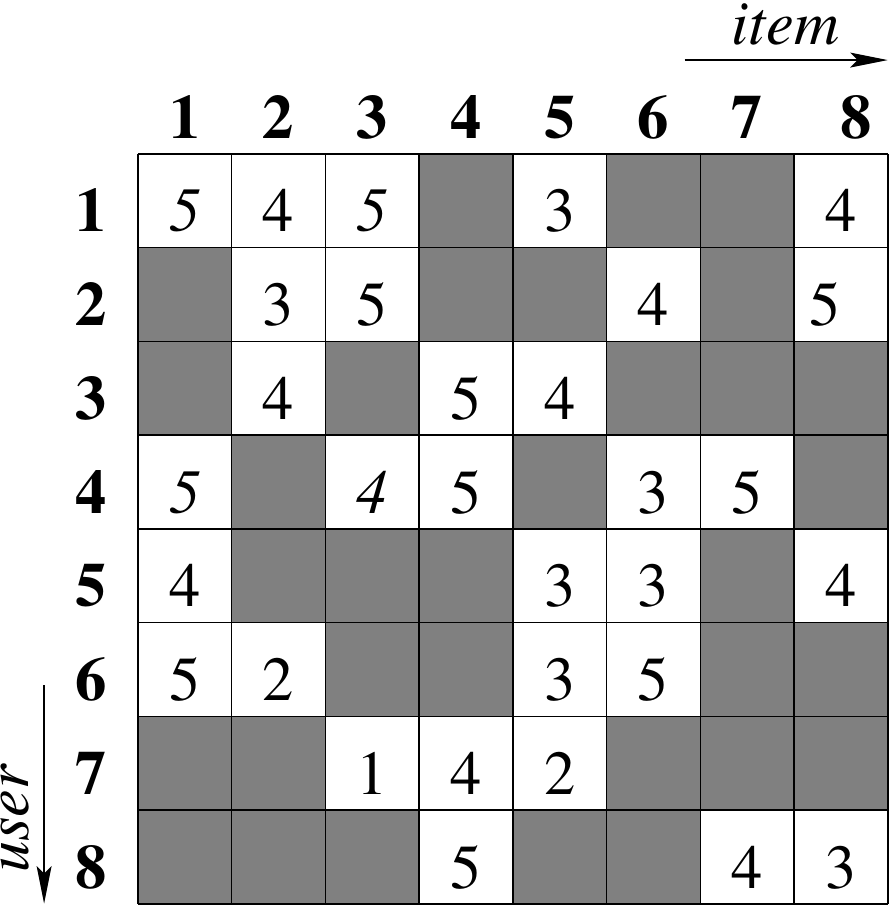}
\end{tabular}
\end{center}
\caption{Preference matrix. Entries are numerical rating (or preference)
			and empty cells are to be filled by the recommender system.}
\label{fig:user-item}
\end{figure}
%

\begin{figure}[htb]
\begin{center}
\begin{tabular}{cc}
\includegraphics[width=0.45\linewidth]{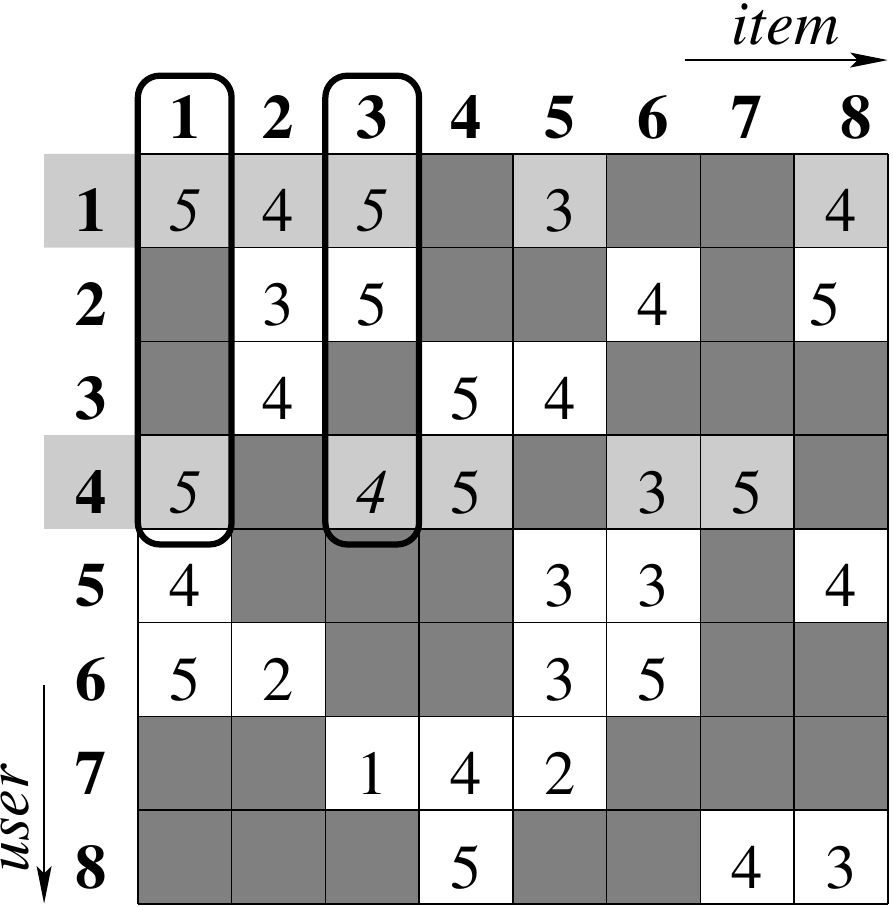} &
\includegraphics[width=0.45\linewidth]{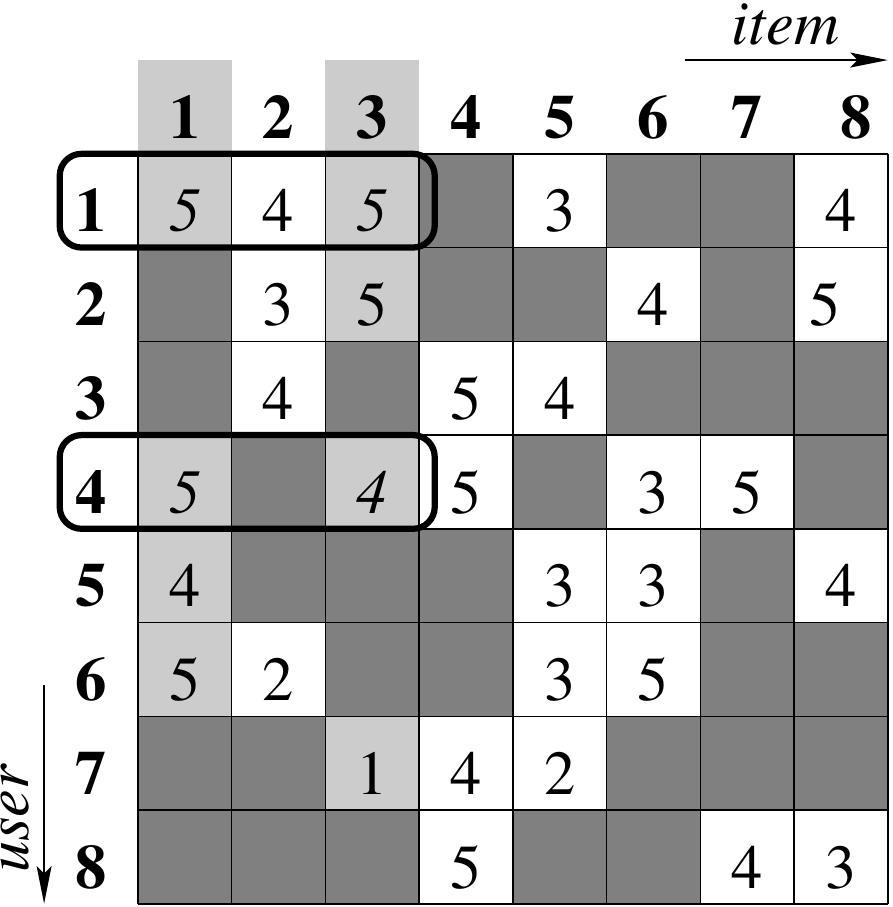}\\
(a) & (b)
\end{tabular}
\end{center}
\caption{User-based correlation (a) and Item-based
			correlation (b).}
\label{fig:PN-correlation}
\end{figure}

This section provides some background on recommender systems
and we refer readers to \cite{adomavicius2005tng} for
a more comprehensive survey.
Let us start with some notations.
Let $\U = \{u_1,\ldots,u_M\}$ be the set of $M$ users (e.g. service subscribers, movie viewers,
Website visitors or product buyers), and $\I = \{i_1,\ldots,i_L\}$ be the set
of $L$ products or items (e.g. services, movies, Webpages or books) that the user can select from. Let us further denote $\M = \{r_{ui}\}$  
the {\em preference matrix} where $u$ is the user index, $i$ is the item index, and $r_{ui}$ is 
the preference or the numerical rating of user $u$ over item $i$
(see Figure~\ref{fig:user-item} for an illustration).
In this paper, we assume that ratings have been appropriately transformed
into integers, i.e. $r_{ui} \in \{1,2,...,S\}$.

Typically, a user usually rates 
only a small number of items and thus making the preference matrix $\M$ extremely sparse.
For example, in the MovieLens dataset that we use in our experiments
(Section~\ref{sec:exp}), only about 6.3\% entries in the $\M$ matrix are filled,
and in large e-commerce sites, the sparsity can be as small as 0.001\%. 
The rating prediction task in recommender systems can be considered as filling
the empty cells in the preference matrix. Of course, due to the data sparsity,
filling all the cells is impractical and often unnecessary because
each user will be interested in a very small set of items. 
Rather, it is only appropriate for a limited
set of entries in each row (corresponding to a user).
Identifying the most relevant entries and ranking them
are the goal of top-$N$ recommendation.

Recommender techniques often fall into three groups:
\emph{content-based}, \emph{collaborative filtering}, and
\emph{hybrid methods} that combines the former two groups.
\\
\\
\textbf{Content-based methods} rely on the content
of items that match a user's profile to make recommendation
using some classification techniques (e.g. see~\cite{mooney2000cbb}).
The content of an item is often referred to the set of attributes
that characterise it. For example, in movie recommendation,
item attributes include movie genres, release date,
leading actor/actress, director, ratings by critics,
financial aspects, movie description and reviews. 
Similarly, user attributes include static information such as 
age\footnote{Strictly speaking,
age is not truly static, but it changes really slowly as long as selling is concerned.}, sex, 
location, language, occupation and marriage status and dynamic 
information such as watching time (day/night/late night),
context of use (e.g. home/theater/family/dating/group/company),
and in case of on-demand videos, what other TV 
channels are showing, what the person has been watching in the past
hours, days or weeks.
\\
\\
\textbf{Collaborative filtering} takes a different approach
in that recommendation is based not only on the usage history
of the user but also on experience and wisdom of 
related people in the user-item network.  Most existing algorithms 
taking some measure of correlation between
co-rating users or co-rated items. One family, known as user-based 
(sometimes memory-based) methods \cite{resnick94grouplens}, 
predicts a new rating of an item based on 
existing ratings on the same item by other users:
\begin{eqnarray}
	r_{ui} = \bar{r}_u + \frac{\sum_{v \in U(i)} s(u,v)(r_{ui} - \bar{r}_v)}{\sum_{v \in U(i)}|s(u,v)|}\nonumber
\end{eqnarray}
where $s(u,v)$ is the similarity between user $u$ and user $v$,
$U(i)$ is the set of all users who rate item $i$, and $\bar{r}_u$ is
the average rating by user $u$.
The similarity $s(u,v)$ is typically measured using Pearson's correlation:
\begin{eqnarray}
	\label{user-cor}
	 \frac{\sum_{i \in I(u,v)}(r_{ui} - \bar{r}_u)(r_{vi} - \bar{r}_v)}
						{\left[\sum_{i \in I(u,v)}(r_{ui} - \bar{r}_u)^2\right]^{\frac{1}{2}}
							\left[\sum_{j \in I(u,v)}(r_{vj} - \bar{r}_v)^2\right]^{\frac{1}{2}}}\nonumber
\end{eqnarray}
where $I(u,v)$ is the set of all items co-rated by users $u$ and $v$.
See Figure~\ref{fig:PN-correlation}a for illustration.
This similarity is computed offline for every pair of users who co-rate
at least one common item.

%
%
The main drawback of user-based methods is in its lack of
efficiency at prediction time because each prediction
require searching and summing over all users who rate
the current item. The set of such users is often
very large for popular items, sometimes including all
users in the database. In contrast, each user typically
rates only a very limited number of items. Item-based methods
\cite{sarwar2001ibc} exploit that fact by simply exchanging the role
of user and item in the user-based approach.
Similarity between items $s(i,j)$ can be computed in several ways
including the (adjusted) cosine between two item vectors, 
and the Pearson correlation. For example, the adjusted cosine
similarity is computed as
\begin{eqnarray}
	\label{item-cor}
	\frac{\sum_{u\in U(i,j)}(r_{ui} - \bar{r}_u )(r_{uj} - \bar{r}_u)}
		{\left[{\sum_{u\in U(i,j)}(r_{ui} - \bar{r}_u)^2}\right]^{\frac{1}{2}}
		\left[{\sum_{v\in U(i,j)}(r_{vj} - \bar{r}_v)^2}\right]^{\frac{1}{2}}}\nonumber
\end{eqnarray}
where $U(i,j)$ is the set of all users who co-rate both items $i$ and $j$.
See Figure~\ref{fig:PN-correlation}b for illustration.
The new rating is predicted as
\begin{eqnarray}
	r_{ui} = \bar{r}_i + \frac{\sum_{j \in I(u)} s(i,j)(r_{uj} - \bar{r}_j)}{\sum_{j \in I(u)} |s(i,j)|}\nonumber
\end{eqnarray}
where $I(u)$ is the set of items rated by user $u$.

Many other methods attempt to build a model of training data that then use
the model to perform prediction on unseen data.
One class of methods employ probabilistic 
graphical models such as Bayesian networks \cite{breese1998eap},
dependency networks \cite{heckerman2001dni}, and 
restricted Boltzmann machines \cite{Salakhutdinov-et-alICML07}. 
Our proposed method using Markov networks fall under the
category of undirected graphical models. It resembles dependency networks
in the way that \emph{pseudo-likelihood} \cite{Besag-74} learning is employed, but
dependency networks are generally inconsistent probabilistic models.
In \cite{Salakhutdinov-et-alICML07}, the authors build
a generative Boltzmann machine for each user with hidden variables, while our method
constructs a single discriminative Markov network 
for the whole database of all ratings. 

Much of other probabilistic work attempts to perform
clustering. This is an important technique for reducing
the dimensionality and noise, dealing
with data sparsity and more significantly,
discovering latent structures. Here the latent structures are
either communities of users with similar tastes
or categories of items with similar features. Some representative
techniques are mixture models, probabilistic latent semantic analysis (pLSA)
\cite{hofmann2004lsm} and latent Dirichlet allocation (LDA)
\cite{marlin2004mur}. These methods try to uncover some
hidden process which is assumed to generate items, users and ratings. In our approach,
no such generation is assumed and ratings are modeled conditionally
given items and users and prior knowledge.


%
Statistical machine learning 
techniques \cite{billsus1998lci,basu1998rcu,zhang2002rsu,basilico2004uca,zitnick2004mec}
have also been used to some extent. 
One of the key observations made is that there is some similarity
between text classification 
and rating prediction \cite{zhang2002rsu}.
However, the main difficulty is that the features in collaborative filtering
are not rich and the nature of prediction is different.
There are two ways to convert collaborative filtering
into a classification problem \cite{billsus1998lci}. The first
is to build a model for each item, and ratings by
different users are treated as training instances. The other
builds a model for each user, and ratings on
different items by this user are considered as
training instances \cite{breese1998eap}. These treatments, however,
are complementary, and thus, there should be a better way
to systematically unify them \cite{basu1998rcu,basilico2004uca}.
That is, the pairs (user,item) are now as independent
training instances.
Our approach, on the other hand, considers the pair
as just a node in the network, thus relaxing
the independence assumption.
\\
\\
\textbf{Hybrid methods} exploit the fact that content-based and collaborative filtering
methods are complementary \cite{balabanovic1997fcb,basu1998rcu,pazzani1999fcc,schein2002mam,basilico2004uca}.
For example, the content-based methods do not suffer from
the so-called \emph{cold-start} problem \cite{schein2002mam} in standard collaborative filtering.
The situation is when new user and new item are introduced to the database, as
no previous ratings are available, purely correlation-based methods cannot work.
On the other hand, content information available
is sometimes very limited to basic attributes that are shared by many items or users.
Prediction by pure content-based methods in that case cannot be personalised
and may be inaccurate. Some work approaches the problem by making independent 
predictions separately using a content-based method and a collaborative filtering
method and then combining the results \cite{claypool1999ccb}. 
Others (e.g. \cite{basilico2004uca}) create joint representation of 
content and collaborative features. We follow the latter approach.

\input{PN.tex}

\section{Experiments}
\label{sec:exp}
In this section, we evaluate our Preference Network 
against well-established correlation methods on the movie
recommendation tasks, which include
rate prediction and top-$N$ item recommendation.

\subsection{Data and Experimental Setup}
We use the MovieLens data\footnote{http://www.grouplens.org},
collected by the GroupLens Research Project at the University of Minnesota
from September 19th, 1997 through April 22nd, 1998.
We use the dataset of 100,000 ratings in the 1-5 scale. This has 943 users and 1682 movies. 
The data is divided into a training set of 80,000 ratings, and the test
set of 20,000 ratings. The training data accounts for 852,848 and 411,546 user-based
item-based correlation features.

We transform the content attributes into a vector of binary indicators.
Some attributes such as sex are categorical and thus
are dimensions in the vector. Age requires
some segmentation into intervals: under 18, 18-24, 25-34, 35-44, 45-49,
50-55, and 56+. We limit user attributes to age, sex and 20 job categories
\footnote{Job list:
administrator, artist, doctor, educator, engineer, entertainment, executive, healthcare,
homemaker, lawyer, librarian, marketing, none, other, programmer, retired, salesman,
scientist, student, technician, writer.}, 
and item attributes to 19 film 
genres
\footnote{Film genres: unknown, action, adventure, animation,
children, comedy, crime, documentary, drama, fantasy,
film-noir, horror, musical, mystery, romance, sci-fi, thriller, war, western.}.
Much richer movie content can be obtained from the Internet Movie 
Database (IMDB)\footnote{http://us.imdb.com}.

%
\subsection{Accuracy of Rating Prediction}
\begin{figure}[htb]
\begin{center}
\begin{tabular}{c}
\includegraphics[width=0.58\linewidth]{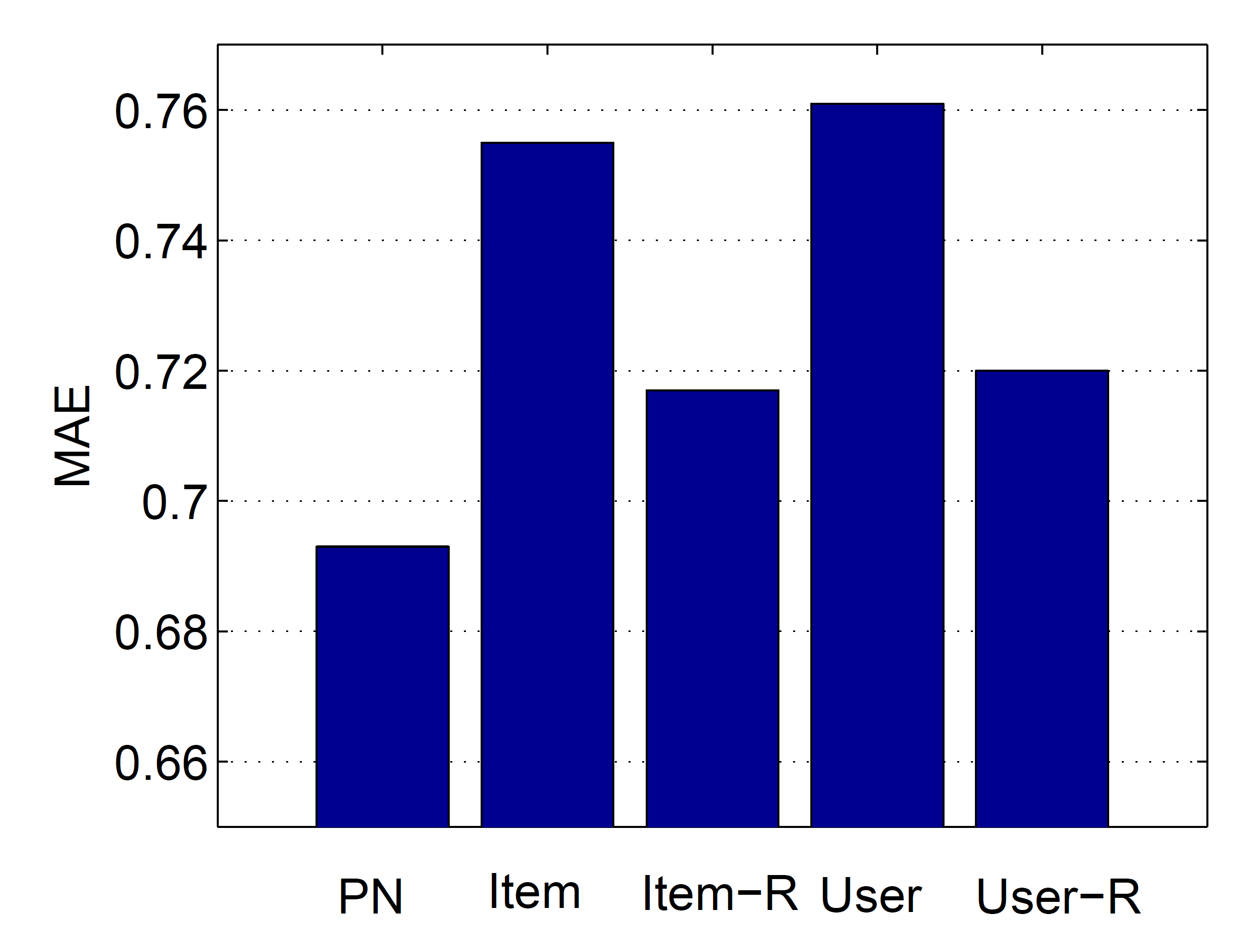}
\end{tabular}
\end{center}
\caption{The mean absolute error of recommendation methods (Item: item-based method,
			and Item-R: item-based method with rounding).}
\label{fig:movilens-small-MAE-round}
\end{figure}

\begin{figure}[htb]
\begin{center}
\begin{tabular}{c}
\includegraphics[width=0.58\linewidth]{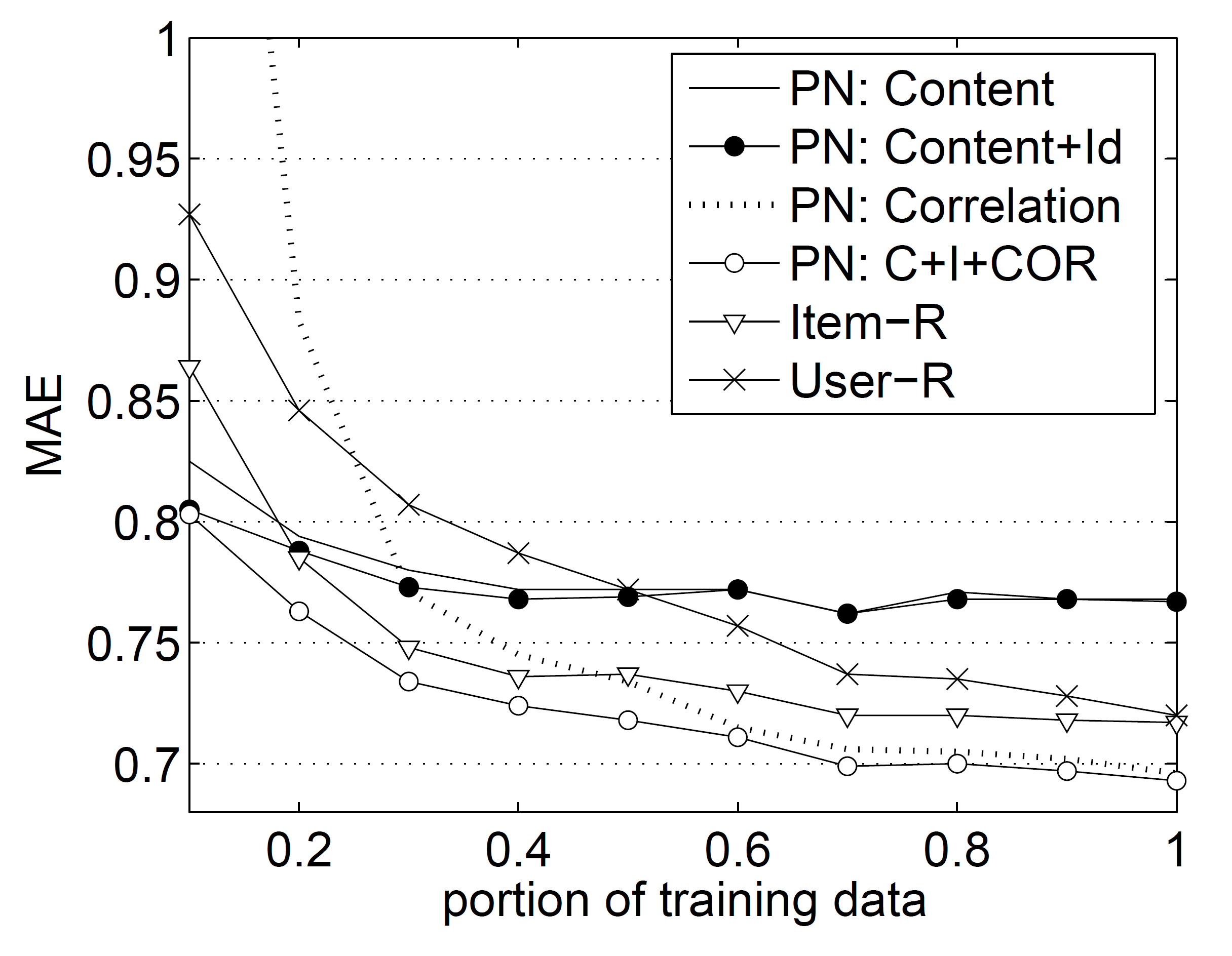} 
\end{tabular}
\end{center}
\caption{The mean absolute error (MAE) of recommendation methods with respect to training
			size of the MovieLens data. (Item: item-based method, and Item-R: item-based method with rounding,
			User: user-based method, User-R: user-based method with rounding,
			Content: PNs with content-based features, C+I+CORR: PNs with 
			content, identity and correlation features).}
\label{fig:movilens-small-MAE}
\end{figure}

\begin{figure}[htb]
\begin{center}
\begin{tabular}{c}
\includegraphics[width=0.58\linewidth]{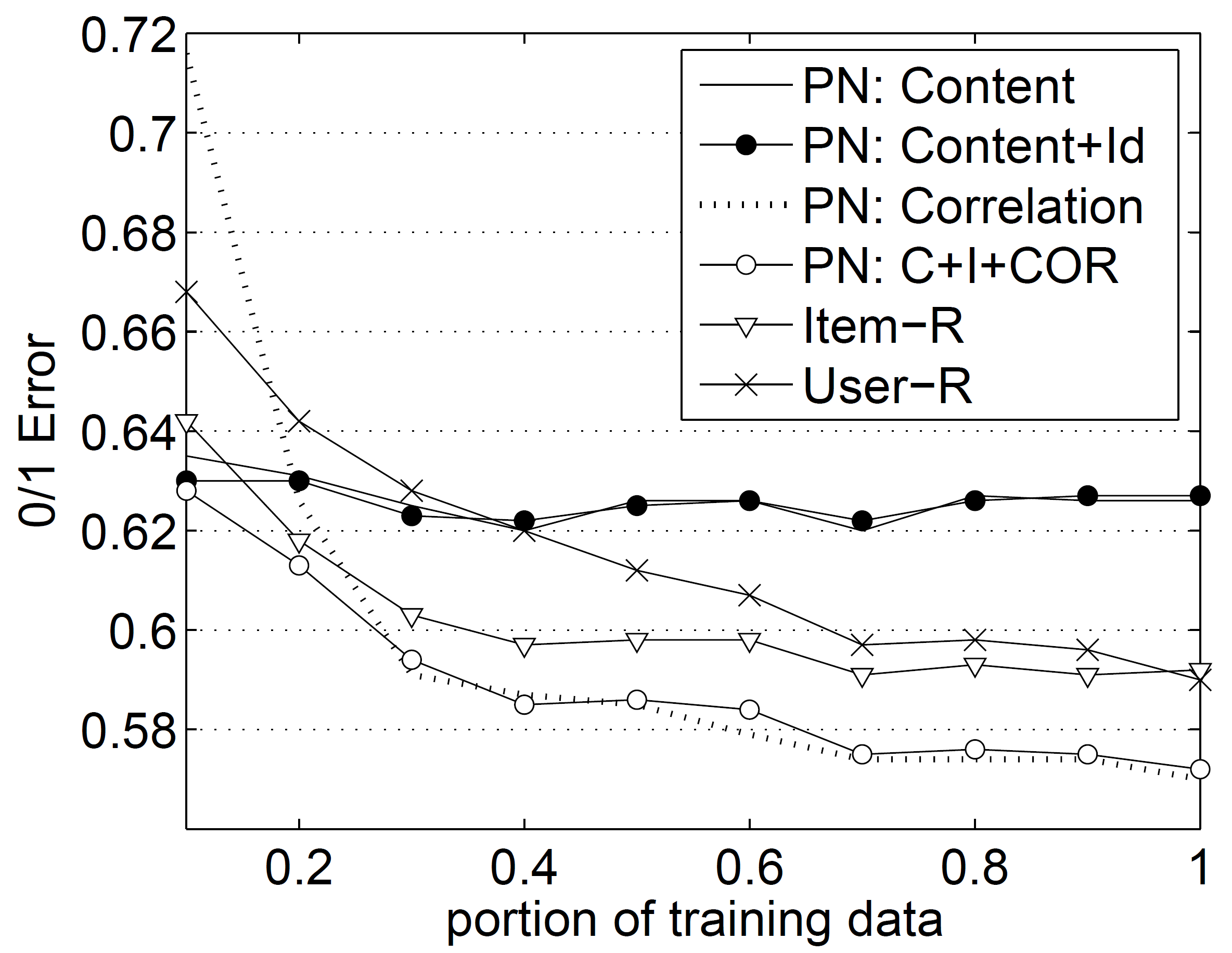}
\end{tabular}
\end{center}
\caption{The mean 0/1 error
			of recommendation methods with respect to training
			size of the MovieLens data. (Item: item-based method, and Item-R: item-based method with rounding,
			User: user-based method, User-R: user-based method with rounding,
			Content: PNs with content-based features, C+I+CORR: Ns with 
			content, identity and correlation features).}
\label{fig:movilens-small-01err}
\end{figure}

In the training phrase, we set the learning rate
$\lambda = 0.001$ and the regularisation term
$\sigma = 1$. We compare our method with well-known user-based \cite{resnick94grouplens} and item-based 
\cite{sarwar2001ibc} techniques (see Section~\ref{sec:recom-sys}).
Two metrics are used: the mean absolute error (MAE) 
\begin{eqnarray}
	\sum_{(u,i) \in \T'} |\hat{r}_{ui} - r_{ui}|/(|\T'|)
\end{eqnarray}
where $\T'$ is the set of rating indexes in the test data,
and the mean 0/1 error 
\begin{eqnarray}
	\sum_{(u,i) \in \T'} \delta(\hat{r}_{ui} \ne r_{ui})/(|\T'|)
\end{eqnarray}
In general, the MAE is more desirable than the 0/1 error because making
exact prediction may not be required and making `closed enough' predictions
is still helpful. As item-based and user-used algorithms output real ratings,
we round the numbers before computing
the errors.  Results shown in Figure~\ref{fig:movilens-small-MAE-round} demonstrate
that the PN outperforms both the item-based and user-based methods.
\\
\\
\textbf{Sensitivity to Data Sparsity}.\\
To evaluate methods against data sparsity, we randomly subsample
the training set, but fix the test set. We report the performance
of different methods using the MAE metric in Figure~\ref{fig:movilens-small-MAE}
and using the mean 0/1 errors in Figure~\ref{fig:movilens-small-01err}.
As expected, the purely content-based method deals with the sparsity in
the user-item rating matrix very well, i.e. when
the training data is limited. However, as the content we use here
is limited to a basic set of attributes, more data does not help the content-based method further. 
The correlation-based method (purely collaborative filtering), 
on the other hand, suffers severely from
the sparsity, but outperforms all other methods when the data is sufficient.
Finally, the hybrid method, which combines all the content, identity and correlation features,
improves the performance of all the component methods, both when data is sparse, and
when it is sufficient.

\subsection{Top-$N$ Recommendation}
We produce a ranked list of items
for each user in the test set so that these items do not appear in the training set.
When a recommended item is in the test set of a user, we call it is a hit.
For evaluation, we employ two measures.
The first is the \emph{expected utility} of the ranked list \cite{breese1998eap},
and the second is the MAE computed over the hits. The expected utility takes into
account of the position $j$ of the hit in the list for each user $u$
\begin{eqnarray}
	R_u = \sum_j \frac{1}{2^{(j-1)/(\alpha-1)}}
\end{eqnarray}
where $\alpha$ is the viewing halflife. Following \cite{breese1998eap}, we
set $\alpha=5$. Finally, the expected utility for
all users in the test set is given as
\begin{eqnarray}
	R = 100\frac{\sum_u R_u}{\sum_u R_u^{max}}
\end{eqnarray}
where $R_u^{max}$ is computed as
\begin{eqnarray}
	R_u^{max} = \sum_{j \in I'(u)}\frac{1}{2^{(j-1)/(\alpha-1)}}
\end{eqnarray}
where $I'(u)$ is the set of items of user $u$ in the test set.

For comparison, we implement a user-based recommendation in that
for each user, we choose 100 best (positively) correlated users
and then rank the item based on the number of times
it is rated by them.
Table~\ref{tab:top-20} reports results of Preference Network
with ranking measure of maximal energy change and expected energy change
in producing the top 20 item recommendations.

\begin{table}[htb]
\begin{center}
\begin{tabular}{|l|c|c|c|} \hline
Method  				& MAE 	& Expected Utility \\ \hline\hline
User-based				& 0.669	& 46.61  \\ \hline
PN (maximal energy)		& 0.603	& 47.43  \\ \hline
PN (expected energy) 	& 0.607 & 48.49  \\ \hline
\end{tabular}
\end{center}
\caption{Performance of top-20 recommendation. PN = Preference Network.}
\label{tab:top-20}
\end{table}

\begin{figure}[htb]
\begin{center}
\begin{tabular}{c}
\includegraphics[width=0.58\linewidth]{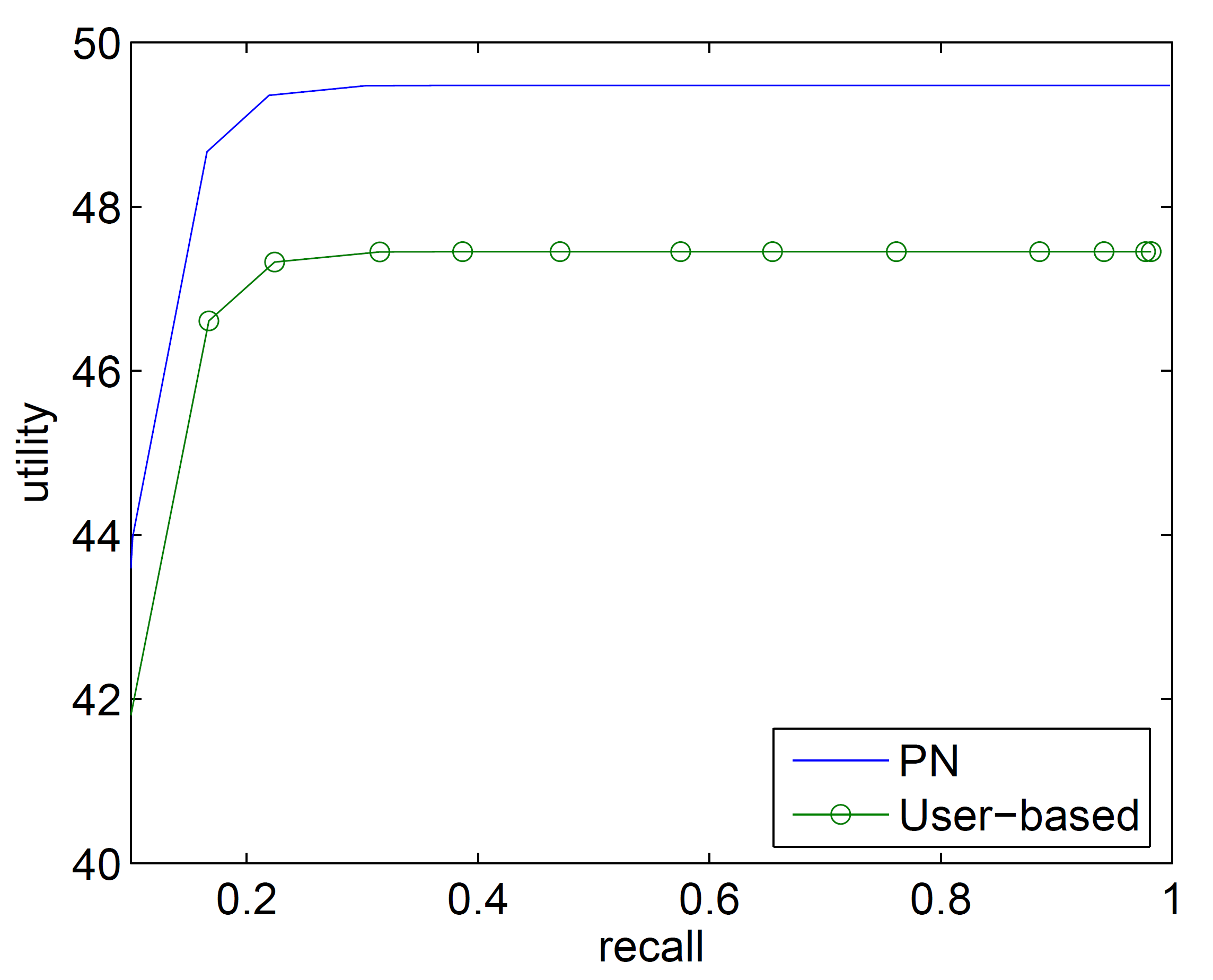}
\end{tabular}
\end{center}
\caption{Expected utility as a function of recall. The larger utility, the better. 
		PN = Preference Network.}
\label{fig:movilens-small-utility-recall}
\end{figure}

\begin{figure}[htb]
\begin{center}
\begin{tabular}{c}
\includegraphics[width=0.58\linewidth]{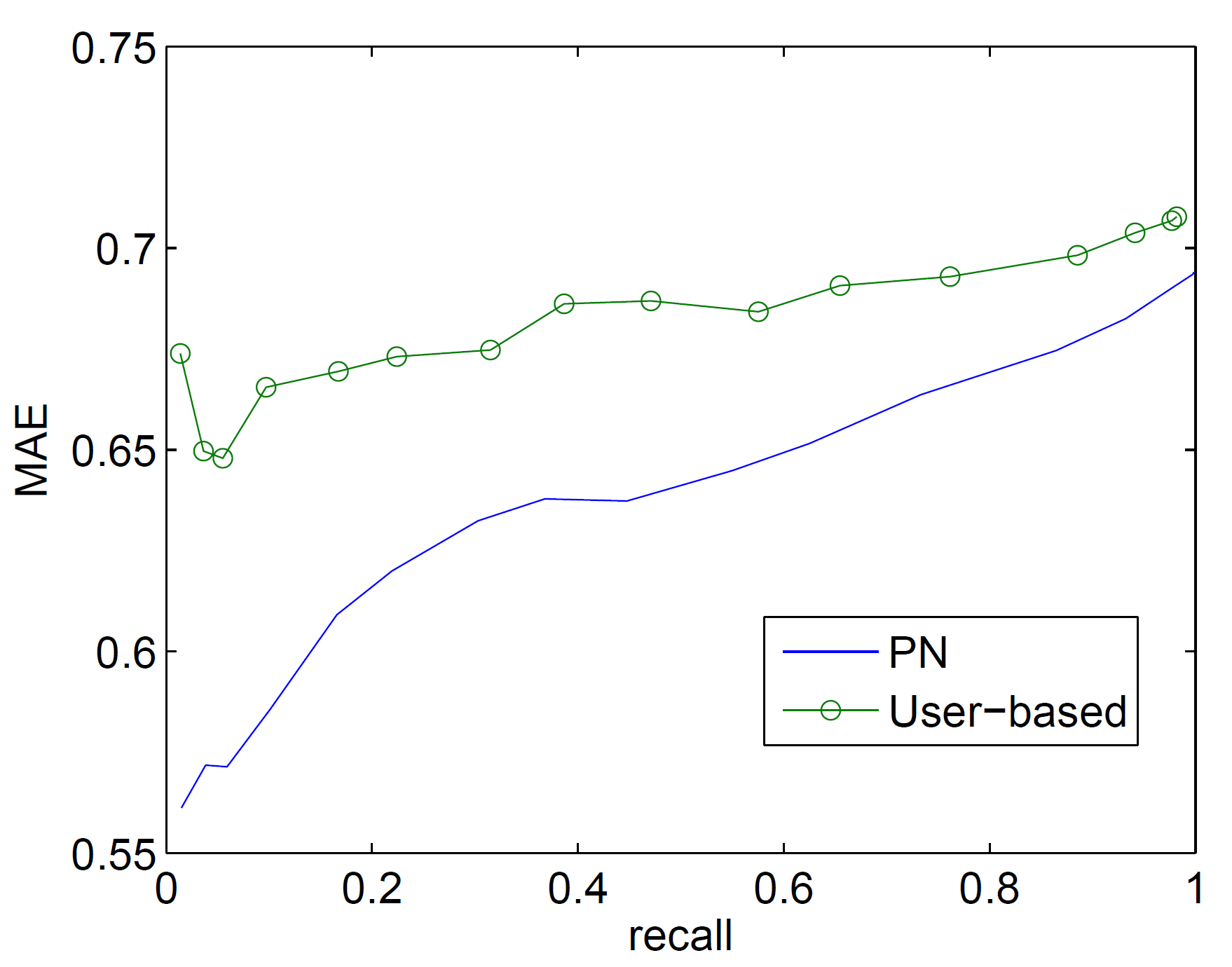}
\end{tabular}
\end{center}
\caption{MAE as a function of recall. The smaller MAE, the better.
			PN = Preference Network.}
\label{fig:movilens-small-MAE-recall}
\end{figure}

We vary the rate of recall by varying the value of $N$, i.e. the recall
rate typically improves as $N$ increases. We are interested in
how the expected utility and the MAE changes as a function of recall.
The expected energy change is used as the ranking criteria for the
Preference Network. Figure~\ref{fig:movilens-small-utility-recall} shows that 
the utility increases as a function of recall rate and reaches
a saturation level at some point. Figure~\ref{fig:movilens-small-MAE-recall}
exhibits a similar trend. It supports the argument that when the recall rate is smaller
(i.e. $N$ is small), we have more confidence on the recommendation. 
For both measures, it is evident that the Preference Network
has an advantage over the user-based method.

\section{Discussion and Conclusions}
\label{sec:discuss}
We have presented a novel hybrid recommendation framework called Preference Networks
that integrates different sources of content (content-based filtering)
and user's preferences (collaborative filtering) into
a single network, combining advantages of both approaches, whilst overcoming shortcomings of 
individual approaches such as the cold-start problem of the collaborative filtering.
Our framework, based on the conditional Markov random fields, are formal to characterise
and amenable to inference. Our experiments show that PNs are competitive
against both the well-known item-based and user-based collaborative
filtering methods in the rating prediction task, and against
the user-based method in the top-$N$ recommendation task. 

Once learned, the PN is a probabilistic database
that allows interesting queries. For example, the set of most influential
items for a particular demographic user group can be identified
based on the corresponding energies. 
Moreover, the conditional nature of the PN supports
fusion of varieties of information into the model through weighted feature functions. 
For example, the features can capture the assertion that if two people
are friends, they are more likely to have similar tastes even though they have not explicitly
provided any common preferences\footnote{Friends are a influential factor
of consumer behaviour via the `word-of-mouth' process}.

Finally, one main drawback the PNs inherit from the user-based methods
is that it may be expensive at prediction time, because it takes into account
 all users who are related to the current one. On-going work will investigate
clustering techniques to reduce the number of pair-wise connections between users.

\appendix
\section{Markov Property and Learning Log-linear Models}
\label{appdx}
This paper exploits an important aspect of Markov networks known as
Markov property that greatly simplifies the computation. Basically, the
property ensures the conditional independence of a variable $r_t$ with
respect to other variables in the network given its neighbourhood
\begin{eqnarray}
	P(r_t|x\backslash r_t,\obs) = P(r_t|\mathcal{N}(t),\obs)
\end{eqnarray}
where $\mathcal{N}(t)$ is the neighbourhood of $r_t$. This explains
why we just need to include the neighbourhood in the Equation~\ref{prediction}. 
This is important because $P(r_t|\mathcal{N}(t),\obs)$ can be easily
evaluated
\begin{eqnarray}
	P(r_t|\mathcal{N}(t),\obs) = 
		\frac{1}{Z_t}\psi_t(r_t,\obs)\prod_{t' \in \mathcal{N}(t)}\psi_{t,t'}(r_t,r_{t'},\obs) \nonumber
\end{eqnarray}
where $Z_t = \sum_{r_t}\psi_t(r_t,\obs)\prod_{t' \in \mathcal{N}(t)}\psi_{t,t'}(r_t,r_{t'},\obs)$.

The parameter update rule in Equation~\ref{param-update} requires
the computation of the gradient of the regularised log pseudo-likelihood
in Equation~\ref{log-ll}, and thus, the gradient
of the log pseudo-likelihood $L = \log P(r_t|\mathcal{N}(t),\obs)$.
Given the log-linear parameterisation in Equations~\ref{potential1} and \ref{potential2},
we have
\begin{eqnarray}
	\frac{\partial \log L}{\partial {\w}_v} &=&
		{\f}_v(r_t,\obs) - \sum_{r'_t}P(r'_t|\mathcal{N}(t),\obs){\f}_v(r'_t,\obs) \nonumber\\
	\frac{\partial \log L}{\partial {\w}_e} &=& 
		{\f}_e(r_t,r'_t,\obs) - \sum_{r'_t}P(r'_t|\mathcal{N}(t),\obs){\f}_e(r'_t,r_{t'},\obs) \nonumber
\end{eqnarray}

\end{document}

%% file: PN.tex
\section{Preference Networks for Hybrid Recommendation}
\label{sec:model}
\newcommand{\obs}{\mathbf{o}}
\newcommand{\vertex}{\mathcal{V}}
\newcommand{\edges}{\mathcal{E}}
\newcommand{\allr}{X}
\newcommand{\C}{\mathcal{C}}
\newcommand{\G}{\mathcal{G}}
\newcommand{\clique}{{\bf r}}
\newcommand{\T}{\mathcal{T}}
\newcommand{\w}{\mathbf{w}}
\newcommand{\f}{\mathbf{f}}
\subsection{Model Description}
\label{model-def}
Let us start with the preference matrix $\M = \{r_{ui}\}$ discussed previously (cf. Sec. 2),
where we treat each entry $r_{ui}$ in $\M$ as a random variable, and thus ideally we
would be interested in a single joint model over $KM$ variables for both the learning
phase and the prediction/recommendation phase. However, in practice, $KM$ is
extremely large (e.g., $10^6 \times 10^6$) making computation intractable.
In addition, such a modeling is unncessary, because, as we have mentioned 
earlier in Section 2, a user 
is often interested in a moderate number of items. 
As a result, we adopt a two-step strategy.
During the learning phase, we limit to model the joint distribution over existing
ratings. And then during the prediction/recommendation phase,
we extend the model to incoporate to-be-predicted entries.

\begin{figure}[htb]
\begin{center}
\begin{tabular}{c}
\includegraphics[width=0.60\linewidth]{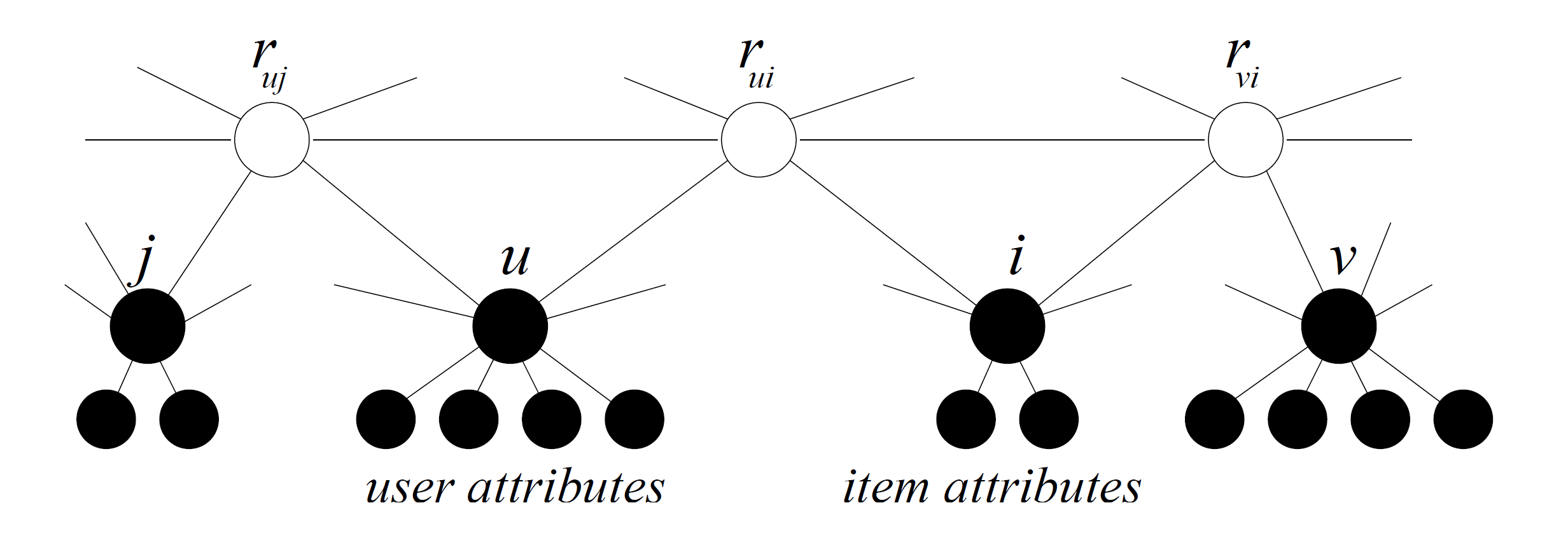}
\end{tabular}
\end{center}
\caption{A fragment of the Preference Network.}
\label{fig:PN-PN}
\end{figure}

We build the model by first representing the ratings and their 
relations using an undirected graph and then defining
a joint distribution over the graph.
Denote by $\G = (\vertex,\edges)$ an undirected graph that has a 
set of vertexes $\vertex$ and a set of edges $\edges$. 
Each vertex in $\vertex$
in this case represents a rating $r_{ui}$ of user $u$ over item $i$ and 
each edge in $\edges$ capture a relation between two ratings. 
The set $\edges$ defines a topological structure
for the network, and specify how ratings are related. 

We define the edges as follows. There is an edge between any two ratings by 
the same user, and an edge between two ratings on the same item.
As a result, a vertex of $r_{ui}$ will be connected with
$U(i)+I(u)-2$ other vertices. Thus, for each user, there is a fully connected 
subnetwork of all ratings she has made, plus
connections to ratings by other users on these items. Likewise,
for each item, there is a fully connected subnetwork of all ratings by different
users on this item, plus connections to ratings on other items by
these users.  The resulting network $\G$ is typically
very densely connected because $U(i)$ can be potentially very large
(e.g. $10^6$). 

Let us now specify the probabilistic modeling of the ratings
and their relations that respect the graph $\G$.
Denote $t = (u,i)$ and let $\T = \{t\}$ be the set of a pair index
(user, item), which corresponds to entries used in each phase.
For notation convenience let $\allr =
\{r_{ui} \mid (u,i) \in \T\}$  denote the joint set of all variables, and the
term `preference' and `rating' will be used exchangeably. When there is no confusion, we use
$r_u$ to denote ratings related to user $u$ and $r_i$ denotes ratings related to item $i$. 

In our approach to the hybrid recommendation task, we
consider attributes of items $\{\mathbf{a}_i\}_{i=1}^L$, 
and attributes of users $\{\mathbf{a}_u\}_{i=u}^M$.
Let $\obs = \{\{\mathbf{a}_i\}_{i=1}^L,\{\mathbf{a}_u\}_{i=u}^M\}$, we 
are interested in modeling the conditional distribution 
$P(\allr|\obs)$ of all user ratings $\allr$ given $\obs$. We employ
the conditional Markov random field~\cite{lafferty01conditional} as
the underlying inference machinery. As $\allr$ collectively represents
users' preferences, we refer this model as {\em Preference Network}.

Preference Network (PN) is thus a conditional Markov random field 
that defines a distribution $P(\allr|\obs)$ 
over the graph $\mathcal{G}$: 
\begin{eqnarray}
	\label{Hammersley-Clifford}
	P(\allr|\obs) &=& \frac{1}{Z(\obs)}\Psi(\allr,\obs), \quad \mbox{where}\nonumber\\
	\Psi(\allr,\obs) &=& \prod_{t \in \vertex}\psi_t(r_t,\obs)
						\prod_{(t,t') \in \edges}\psi_{t,t'}(r_t,r_{t'},\obs)
\end{eqnarray}
where $Z(\obs)$ is the normalisation constant
to ensure that $\sum_{\allr}P(\allr|\obs)=1$, and $\psi(.)$ is
a positive function, often known as \em potential\em.
More specifically, $\psi_t(r_t,\obs)$ encodes the content information
associated with the rating $r_t$ including the attributes
of the user and the item. On the other hand,
$\psi_{t,t'}(r_t,r_{t'},\obs)$ captures the correlations between
two ratings $r_t$ and $r_{t'}$. Essentially, when there are
no correlation potentials, the model is purely content-based, and
when there are no content potentials, 
the model is purely collborative-filtering.
Thus the PN integrates both types of recommendation in a seamlessly
unified framework.

The contribution of content and correlation potentials to the joint
distribution will be adjusted by weighting parameters associated with them.
Specifically, the parameters are encoded in potentials as follows
\begin{eqnarray}
	\label{potential1}
	\psi_t(r_t,\obs) &=& \exp\left\{\w_{v}^{\top}\f_v(r_t,\obs)\right\} \\
	\label{potential2}
	\psi_{t,t'}(r_t,r_{t'},\obs) &=& \exp\left\{\w_{e}^{\top}\f_e(r_t,r_{t'},\obs)\right\}
\end{eqnarray}
where $\f(.)$ is the feature vector and $\w$ is the corresponding weight
vector. Thus together with their weights, the features realise the contribution of
the content and the strength of correlations between items and users. 
The design of features will be elaborated further in Section~\ref{feature}.
Parameter estimation is described in Section~\ref{learning}.

\subsection{Feature Design and Selection}
\label{feature}

Corresponding to the potentials in Equations~\ref{potential1} and \ref{potential2},
there are attribute-based features and correlation-based features.
Attribute-based features include user/item identities and contents.

\textbf{Identity Features}. Assume that the ratings are integer, ranging from 1 to $S$.
We know from the database the average rating $\bar{r}_i$ of item $i$
which roughly indicates the general quality of the item with respect to those who have rated it.
Similarly, the average rating $\bar{r}_u$ by user $u$
over items she has rated roughly indicates the
user-specific scale of the rating because the same rating of $4$ may mean 
`OK' for a regular user, but may mean `excellent' for a critic.
We use two features {\em item-specific} $f_i(r_{ui},i)$ and {\em
user-specific} $f_u(r_{ui},u)$:
\[
f_i(r_{ui},i) = g(|r_{ui} - \bar{r}_i|), \quad f_u(r_{ui},u) = g(|r_{ui} - \bar{r}_u|)
\]
where $g(\alpha) = 1 - \alpha/(S-1)$ is used to ensure that the feature values is normalized 
to $[0,1]$, and when $\alpha$ plays the role of rating deviation, $g(\alpha) = 1$ for $\alpha=0$.

\textbf{Content Features}. For each rating by user $u$ on item $i$, we have a set of item attributes 
$\mathbf{a}_i$ and set of user attributes $\mathbf{a}_u$.  
Mapping from item attributes to user preference can be carried out through
the following feature
\begin{eqnarray}
	\f_u(r_{ui}) = \mathbf{a}_i g(|r_{ui} - \bar{r}_u|) \nonumber
\end{eqnarray}
Similarly, we are also interested in seeing the classes of users
who like a given item through the following mapping
\begin{eqnarray}
	\f_i(r_{ui}) = \mathbf{a}_u g(|r_{ui} - \bar{r}_i|) \nonumber
\end{eqnarray}

\textbf{Correlation Features}. We design two features to capture correlations between 
items or users. Specifically, the {\em item-item}
$f_{i,j}(\cdot)$ features capture the fact that if a user rates two items then after 
offsetting the goodness of each item, the ratings may be similar
\begin{eqnarray}
	f_{i,j}(r_{ui},r_{uj}) &=& g(|(r_{ui}-\bar{r}_i) - (r_{uj}-\bar{r}_j)|) \nonumber 
\end{eqnarray}
Likewise, the {\em user-user} $f_{u,v}(\cdot)$ features capture
the idea that if two users rate the same item then the ratings, after offset by user's own
scale, should be similar: 
\begin{eqnarray}
	f_{u,v}(r_{ui},r_{vi}) &=& g(|(r_{ui}-\bar{r}_u) - (r_{vi}-\bar{r}_v)|) \nonumber
\end{eqnarray}
Since the number of correlation features can be large, making
model estimation less robust, we select only item-item features
with positive correlation (given in Equation~\ref{item-cor}), and
user-user features with positive correlations (given in Equation~\ref{user-cor}).

\subsection{Parameter Estimation}
\label{learning}
Since the network is densely connected, learning methods based on
the standard log-likelihood $\log P(\allr|\obs)$ are not applicable. 
This is because underlying inference for computing the log-likelihood
and its gradient is only tractable for simple networks with simple chain
or tree structures~\cite{Pearl88}. As a result, we resort to the simple but effective
\em pseudo-likelihood \em learning method \cite{Besag-74}. 
Specifically, we replace the log likelihood
by the regularised sum of log local likelihoods
\begin{eqnarray}
	\label{log-ll}
	\mathcal{L}(\w) = \sum_{(u,i)\in \T} \log P(r_{ui}|\mathcal{N}(u,i),\obs) - \frac{1}{2}\bar{\w}^{\top}\bar{\w}
\end{eqnarray}
where, $\mathcal{N}(u,i)$ is the set of neighbour ratings that are connected
to $r_{ui}$. As we mentioned earlier, the size of
the neighbourhood is $|\mathcal{N}(u,i)| = U(i) + I(u) - 2$.
In the second term in the RHS,
$\bar{\w} = \w/\boldsymbol{\sigma}$ (element-wise division, 
regularised by a prior diagonal
Gaussian of mean $0$ and standard deviation vector $\boldsymbol{\sigma}$).

Finally, the parameters are estimated by maximising the pseudo-likelihood
\begin{eqnarray}
	\hat{\w} = \arg\max_{\w} \mathcal{L}(\w)
\end{eqnarray}

Not only is this regularised pseudo-likelihood simple to implement,
it makes sense since the local conditional distribution 
$P(r_{ui}|\mathcal{N}(u,i),\obs)$ is used in prediction (Equation~\ref{prediction}).
We limit ourselves to supervised learning in that
all the ratings $\{r_{ui}\}$ in the training data are known.
Thus, $\mathcal{L}(\w)$ is a concave function
of $\w$, and thus has a unique maximum. 

To optimise the parameters, we use a simple stochastic
gradient ascent procedure that updates the parameters
after passing through a set of ratings by each user:
\begin{eqnarray}
	\label{param-update}
	\w_u \leftarrow \w_u + \lambda \nabla \mathcal{L}(\w_u)
\end{eqnarray}
where $\w_u$ is the subset of parameters that are
associated with ratings by user $u$, and $\lambda > 0$ is the 
learning rate. 
Typically, 2-3 passes through the 
entire data are often enough in our experiments.
Further details of the computation are included in
Appendix~\ref{appdx}.


\subsection{Preference Prediction}
Recall from Section~\ref{model-def} that we employ a two-step modeling.
In the learning phase (Section~\ref{learning}), 
the model includes all previous ratings. Once the model has been estimated, we
extend the graph structure to include
new ratings that need to be predicted or recommended. 
Since the number of ratings newly added is typically small
compared to the size of existing ratings, it can be assumed that
the model parameters do not change.

The prediction of the rating $r_{ui}$ for user $u$ over item $i$ is 
given as
\begin{eqnarray}
	\label{prediction}
	\hat{r}_{ui} = \arg\max_{r_{ui}}P(r_{ui} \mid \mathcal{N}(u,i),\obs)
\end{eqnarray}
The probability $P(\hat{r}_{ui}|\mathcal{N}(r_{ui}),\obs)$ is the measure of 
the \emph{confidence} or ranking level in making this prediction. This can be useful in
practical situations when we need high precision, that is, only ratings with the 
confidence above a certain threshold are presented to the users.

We can jointly infer the ratings $r_u$ of given user $u$ on a subset of items 
$\mathbf{i} = (i_1,i_2,..)$ as follows
\begin{eqnarray}
	\label{joint-predict-user}
	\hat{r}_u = \arg\max_{r_u}P(r_u \mid \mathcal{N}(u),\obs)
\end{eqnarray}
where $\mathcal{N}(u)$ is the set of all existing ratings
that share the common cliques with ratings by user $u$.
In another scenario, we may want to recommend 
a relatively new item $i$ to a set of promising users, we can 
make joint predictions $r_i$ as follows
\begin{eqnarray}
	\label{joint-predict-item}	
	\hat{r}_i = \arg\max_{r_i}P(r_i \mid \mathcal{N}(i),\obs)
\end{eqnarray}
where $\mathcal{N}(i)$ is the set of all existing ratings
that share the common cliques with ratings of item $i$.
It may appear non-obvious that a prediction may depend
on unknown ratings (other predictions to be made) but this is the advantage
of the Markov networks. However, joint predictions for a user are only possible
if the subset of items is small (e.g. less than 20) because
we have a completely connected subnetwork for this user.
This is even worse for joint prediction of an item because
the target set of users is usually very large. 

\subsection{Top-$N$ recommendation}
In order to provide a list of top-$N$ items
to a given user, the first step is usually to identify a
candidate set of $C$ promising items, where $C \ge N$. 
Then in the second step, we rank and choose the best $N$ 
items from this candidate set according to some measure of relevance.
\\
\\
\textbf{Identifying the candicate set}.\\
This step should be as efficient as possible and $C$ should be relatively small
compared to the number of items in the database. 
There are two common techniques used in user-based and
item-based methods, respectively. In the user-based technique,
first we idenfify a set of $K$ most similar
users, and then take the union of all items co-rated
by these $K$ users. Then we remove items that the user has previously rated. 
In the item-based technique
\cite{deshpande2004ibt}, for each item the user has rated,
we select the $K$ best similar items that the user has not rated.
Then we take the union of all of these similar items.

Indeed, if $K \rightarrow \infty$, or equivalently, we use all
similar users and items in the database, then the item sets
returned by the item-based and user-based techniques are 
\emph{identical}. To see why, we show that every candidate $j$
returned by the item-based technique is also the candidate
by the user-based techqnique, and vice versa.
Recall that a pair of items is said to
be similar if they are jointly rated by the same user. 
Let $I(u)$ be the set of items rated by the current user $u$.
So for each item $j \notin I(u)$ similar to
item $i \in I(u)$, there must exist a user $v \ne u$ so
that $i,j \in I(v)$. Since $u$ and $v$ jointly rate
$i$, they are similar users, which mean that $j$ is also
in the candidate set of the user-based method. Analogously, 
for each candidate $j$ rated by user $v$, who is similar to $u$, and $j \notin I(u)$, 
there must be an item $i \ne j$ jointly rated by both $u$ and $v$. Thus $i,j \in I(v)$, and
therefore they are similar. This means that $j$ must be a candidate
by the item-based technique.


In our Preference Networks, the similarity measure is replaced
by the correlation between users or between items. 
The correlation is in turn captured by the corresponding
correlation parameters. Thus, we can use either the user-user correlation
or item-item correlation to identify the candicate set. Furthermore,
we can also use both the correlation types and take the union of the
two candidate sets. 
\\
\\
\textbf{Ranking the candidate set}.\\
The second step in the top-$N$ recommendation is to 
rank these $C$ candicates according to some scoring methods.
Ranking in the user-based methods is often based
on item popularity, i.e. the number of users in the neighbourhood
who have rated the item.  Ranking in the item-based methods \citep{deshpande2004ibt}
is computed by considering not only the number of raters but
the similarity between the items being ranked and the set
of items already rated by the user.

Under our Preference Networks formulation, we propose
to compute the change in system energy and use it
as ranking measure. Our PN can be thought 
as a stochastic physical system whose energy is related
to the conditional distribution as follows
\begin{eqnarray}
	P(\allr|\obs) = \frac{1}{Z(\obs)}\exp(-E(\allr,\obs))
\end{eqnarray}
where $E(\allr,\obs) = - \log \Psi(\allr,\obs)$ is the system energy. 
Thus the lower energy the system state $\allr$ has, the more
probable the system is in that state. Let $t = (u,i)$, from Equations~\ref{potential1} and \ref{potential2}, we can
see that the system energy is the sum of node-based energy and interaction energy
\begin{eqnarray}
	E(\allr,\obs) &=& \sum_{t \in \vertex} E(r_t,\obs) + \sum_{(t,t') \in \edges} E(r_t,r_{t'}\obs)\nonumber
\end{eqnarray}
where
\begin{eqnarray}
	E(r_t,\obs) &=& -{\w}_v^{\top}{\f}_v(r_{t},\obs) \\
	E(r_t,r_{t'},\obs) &=& -{\w}_e^{\top }{\f}_e(r_{t},r_{t'},\obs)
\end{eqnarray}

Recommending a new item $i$ to a given user $u$ is equivalent to extending the system
by adding new rating node $r_{ui}$.  The change
in system energy is therefore the sum of node-based energy of the new node,
and the interation energy between the node and its neighbours.
\begin{eqnarray}
	\Delta E({r}_{t},\obs) = E({r}_{t},\obs)+\sum_{t' \in \mathcal{N}(t)}E({r}_{t},r_{t'},\obs)\nonumber
\end{eqnarray}
For simplicity, we assume that the state of the existing system does not
change after node addition. Typically, we want the extended system to 
be in the most probable state, or equivalently the system state with lowest energy.
This means that the node that causes the most reduction of
system energy will be prefered.  Since we do not know the correct
state $r_t$ of the new node $t$, we may guess by predicting
$\hat{r}_{t}$ using Equation~\ref{prediction}. Let us call the energy reduction
by this method the \emph{maximal energy change}. Alternatively, 
we may compute the \emph{expected energy change} to account
for the uncertainty in the preference prediction
\begin{eqnarray}
	\mathbb{E}[\Delta E(r_t,\obs)] = \sum_{r_{t}} P(r_{t}| \mathcal{N}(t),\obs)\Delta E(r_{t},\obs)
\end{eqnarray}